\documentstyle[aas2pp4,psfig]{article}
\begin{document}

\title{\centerline{GALAXY MORPHOLOGICAL SEGREGATION IN CLUSTERS:}
\centerline{ LOCAL VS. GLOBAL CONDITIONS}}

\author{Mariano Dom\'{\i}nguez, Hern\'an Muriel and Diego G. Lambas}

\affil{Grupo de Investigaciones en Astronom\'{\i}a Te\'orica y Experimental (IATE)}

\affil{Observatorio Astron\'omico de C\'ordoba, Laprida 854, 5000 C\'ordoba, Argentina.\\}

\affil{CONICET, Buenos Aires,Argentina. \\
mardom@oac.uncor.edu, hernan@oac.uncor.edu, dgl@oac.uncor.edu}

\centerline{\bf{ABSTRACT}}

We study the relative fraction of galaxy morphological types 
in clusters, as a function 
of the projected local galaxy density and 
different global parameters: cluster projected gas density, 
cluster projected total mass density , and  reduced clustercentric distance.
Since local and global densities are correlated, we have
considered different tests to search for the parameters 
to which segregation show the strongest dependence. 
Also, we have explored the results of our analysis applied to the central regions 
of the clusters and their outskirts.
We consider  a sample of clusters of galaxies with temperature estimates 
to derive the projected mass density profile and the 500 density contrast
radius ($r_{500}$) using the NFW model and the scaling relation respectively. 
The X-ray surface brightness profiles are used to obtain the 
projected gas density assuming the hydrostatic equilibrium model. 
Our results suggest that the morphological segregation in clusters 
is controlled by the local galaxy density in the 
outskirts. On the other hand, the global projected mass density,
shows the strongest correlation with the fraction of morphological types
in the central high density region, with a marginal dependence on the
local galaxy density.



\section{INTRODUCTION}

The difference between the population of galaxies in the field and in clusters is well known 
since the early 1930s. Oemler (1974) analyzed the proportion of elliptical, S0 and spiral 
galaxies and defined three different types of clusters: spiral rich, spiral poor and cD 
clusters. Melnick and Sargent (1977) showed that the inner region of clusters are 
typically populated by ellipticals and spirals are predominating in the outer regions
 (the well known morphological segregation, MS). 

In a classic work, Dressler 1980b analyzed a sample of 55 nearby clusters 
finding
a correlation between the fraction of different morphological types T 
and the local projected galaxy density $\Sigma$ (hereafter T-$\Sigma$ relation). 
Dressler concludes that 
galaxy morphology is a function of the local clustering rather than 
global conditions related to
the cluster environment.

Samrom$\grave{a}$ and Salvador Sole (1990) have consider a test to explore the nature of the
morphological segregation. Using Dressler's data,
artificial clusters were generated by  
randomly repositioning the polar coordinates of galaxy members
around the center of the clusters. 
In this way  any subclustering present is erased maintaining the
radial clustercentric distance of each galaxy so that the 
global cluster properties are conserved. 
The T-$\Sigma $ relation is analyzed  for the randomized galaxies in the clusters
and compared to the results for the real galaxy positions. The results of these
tests are indistinguishable indicating that the T-$\Sigma $ 
relation is controlled by global conditions rather than local subclustering.  

Whitmore et al. (1991,1993, hereafter WGJ ) re-examined Dressler's sample of galaxies 
in clusters and suggested that the morphology-clustercentric distance relation is more 
fundamental than the morphology-local galaxy density relation. These authors  use the 
clustercentric normalized distance as the independent parameter,
and compare the morphological fractions at the  
same normalized clustercentric radii but with different values of the local galaxy density. 
WGJ use a "characteristic radius", ${r_{opt}}$, as the radius at which the cumulative 
projected galaxy density falls bellow 20 galaxies $Mpc^{ -2}$. These authors find that for 
small radii (about 0.5 Mpc) the elliptical fraction rise rapidly. On the other hand, 
the spiral fraction is essentially zero at the cluster center. 
WGJ interpret these results as an indication that cluster center conditions 
play a  key role in determining galaxy morphologies in clusters and suggest 
that a destructive rather than a formation mechanism
may be controlling their relative fractions. 

The question whether local galaxy density or radial distance from the cluster center is 
best correlated with morphological types is still open. Dressler et al. (1997) reanalyzed 
the morphology-density relation by dividing their sample into centrally 
concentrated, regular clusters; and low-concentration, irregular clusters, finding 
a similar T-$\Sigma$ relation in both subsamples.  Nevertheless, they find a significant 
excess of SO's and a smaller spiral fraction in the centrally 
concentrated clusters when compared to the low-concentration systems.  

Dressler et al. (1997, hereafter D97) also analyzed a sample of 10 clusters at redshifts between 0.37 and 
0.56, and derive the corresponding T-$\Sigma$ relation. For these  distant clusters the 
authors find that the fraction  of S0 galaxies is 2-3 times smaller than in low-redshift 
clusters suggesting that S0's are generated in large numbers only after cluster 
virialization. 

It should be recalled, however, the possible presence of systematic
effects related to projection biases in the selection of clusters
from two dimensional data (see for instance Valotto, Moore, and Lambas, 2001), which
may cause an artificial increase of the late type fraction in distant
clusters.

Several mechanisms have been proposed in order to explain the morphological segregation. 
Among them we can mention ram pressure stripping (Gunn and Gott 1972, Abadi et al. 1999),
 gas evaporation (Cowie and Songaila 1977), merging (Lavery and Henry 1988), galaxy 
harassment (Moore et al. 1996), tidal striping (galaxy-galaxy and from mean cluster field; 
 Bird and Valtonen 1990), tidal shaking 
(galaxy-galaxy and from the mean cluster field; Miller 1988), galactic cannibalism 
(Ostriker and Tremaine 1975), truncated star 
formation (Larson et al. 1980), etc. 
Also, it should be considered the effects provided by the 
initial conditions which may also play a 
role in the morphological segregation. In hierarchical models of structure formation
such as CDM, different scales become non-linear simultaneously, so that the initial conditions
of galaxy formation in cluster and group environments may differ 
from the field (intergalactic medium, tidal effects, mergers, etc).

In this work 
we study the relative fraction of galaxy morphologies in clusters  as a 
function of global cluster parameters (projected mass and gas density profile),
and the local projected galaxy density. 
Our analysis is aimed to provide a better understanding of the 
relevance of the proposed mechanisms  
involved in morphological segregation (MS) in clusters
given their different dependence on gas, mass and galaxy content. 

The paper is organized as follows. In Section 2 we briefly review the data used for this study.
In Section 3 we explore the MS as a function of different global and 
local parameters, and perform a simple test to compare their relative importance.
Section 4 provides a discussion of the  main 
results and some implicances for  galaxy 
evolution in clusters.
We include in Appendix A a short discussion on the corrections applied to deal with
observational biases.
  

\section{DATA}

The data consists of the clusters originally analyzed by Dressler 1980a restricted to
those objects with detected intracluster gas X-ray emission and 
determination of its mean temperature and surface brightness 
distribution.

Dressler 1980a survey provides morphological determination for 
$\sim$ 6000 galaxies  
in 55 nearby clusters of galaxies. From this sample we have
selected a sub-sample of 22 rich clusters  (see Table 1) with estimated
(or measured)  gas  temperature  and X-ray surface brightness information. 
From Jones and Forman (1999) analysis of Einstein X-ray images
we obtain the  central gas density $\rho_0$, 
the  core radii $r_c$ and the $\beta$  parameter for our subsample. 
Temperatures are taken from different sources in the literature and as it can be seen 
in Table 1, our sample comprises a wide range of temperatures going
from poor cluster environments
($\sim$ 1.5 keV) to massive clusters ($\sim$ 8 keV). Cluster center coordinates in
Table 1 correspond to the maximum of the X-ray emission. 
The other parameter quoted in this table are:
Column 8: cluster redshift $z$; column 9: cluster temperature $T_{X}$; 
column 10: core radius $r_c$; 
column 11: $\beta$ parameter;
column 12: central density $\rho_0$;
column 13: optical core radius ${r_{opt}}$, and
column 14: 500 overdensity radius $r_{500}$.

We have excluded clusters with peculiarities in their X-ray 
luminosity distribution such as double 
clusters consisting in two sub-structures of comparable size and luminosity, 
objects with a significant presence of substructure, 
or strong departures from sphericity.


\section{METHODS AND ANALYSIS}

In this section we explore the dependence of the relative fraction of 
galaxy morphological types in clusters as a function of  
the local galaxy environment, namely the projected local galaxy density 
as defined by Dressler 1980b, and different global parameters: 
cluster projected total mass density and gas density, 
and two different reduced clustercentric distances.

\subsection{Local galaxy density}

In order to assess the relevance of local 
processes that may affect galaxy morphology, we compute the relative fraction of
galaxy morphologies as a function of the projected local galaxy density
in the same way as D97. 
We define the local galaxy density $\Sigma_{Gal}$ in the position of each galaxy in 
our sample, using the same procedure as D97. 
We compute the rectangular area that comprises the ten nearest
galaxies around each object and correct for completeness in luminosity and contamination from
projections using the same methods described by WGJ, and 
discussed in Appendix A. 
The results are shown in Figure 1 which show similar results than those
given by Dressler et al. (1997) indicating that the results of our subsample of 
X-ray clusters are representative of optically selected galaxy systems. 
Errors bars in all figures of this paper were
computed using the bootstrap resampling technique (Barrow 1984). 

\subsection{Projected cluster mass density}

The effects of the global cluster environment  
on galaxy morphology requires to compute parameters
that quantify the effects of the cluster as a whole at the position of each galaxy.
The total mass density is an important parameter of clusters that 
should be considered given its relevance to the several processes (eg. cluster tidal field, etc) that could affect galaxy morphology.

Navarro et al. (1995, hereafter NFW) have proposed an analytic universal density 
profile of dark matter halos
based on numerical simulations  and analytical models 
assuming spherical symmetry and accretion onto an initially overdense 
perturbation. NFW fitting function is
 
(1) $$ \rho(x) = \frac{\rho_s}{x(1+x)^2}  ,   x=r/r_{\delta_c}   $$

which describes  cluster mass density profiles, where 
$r_{\delta_c}$ is the radius 
corresponding to a mean over-density $\delta_c$. 
This function has the advantage 
that the dark matter 
halos can be described with a single parameter over a broad rage of halo masses. Based 
on numerical simulations, Evrard et al. (1996) predict that the average cluster 
temperature $T_X$ strongly correlates with $r_{\delta_c}$, and propose the following
relation:

(2) $$ r_{\delta_c} (T_X)=r_{10}(\delta_c)*(T_X / 10 keV) $$

The normalization $r_{10}(\delta_c)$ corresponds to the radial scale of 10 keV
clusters at density contrast $\delta_c$ (2.48 Mpc) and 
we adopt a standard overdensity parameter $\delta_c=500$.

Bartelmann (1996) has derived the projected mass
density $\Sigma_{Mass}$ profile corresponding to the NFW function: 

(3) $$ \Sigma_{Mass}(x) = \frac{2 \rho_s r_{\delta_c}}{x^2-1}*f(x)  $$

$$
\mbox{f}(x)=\left\{
\begin{array}{rl}
1 - \frac{2}{\sqrt{x^2-1}} arctanh \sqrt{\frac{x-1}{x+1}}, &\mbox{ if $x>1$} \\
1 - \frac{2}{\sqrt{1-x^2}} arctanh \sqrt{\frac{1-x}{1+x}}, &\mbox{ if $x<1$} \\
0, & \mbox{ if $x=1$}
\end{array} \right.
$$

We consider the mean cluster temperatures quoted in Table 1 
in equation (2) to derive $R_{500}$.  
Then the projected  mass density $\Sigma_{Mass}$ at the position of each galaxy
can then be obtained by equation (3).

For all clusters we compute the relative 
fraction of morphological types in bins of $\Sigma_{Mass}$.
The results are shown  
in Figure 2 where 
it can be appreciated a very significant dependence of 
the relative  fraction of galaxy morphologies on the local mass density inferred from the
projected NFW profile.

Local vs. global effects might be reflected in the different dependence of the morphological segregation
on  $\Sigma_{Gal}$ and $\Sigma_{Mass}$.
By comparison of figures 1 and 2, it can be appreciated that both
 $\Sigma_{Mass}$ and $\Sigma_{Gal}$ provide a good correlation with the relative fraction of
morphological types. However, given  
the correlation between these two densities as shown 
in Figure 3, it is important to analyze  
whether they are primary or secondary parameters in the morphological segregation.\\ 
In order to address this point, we consider for simplicity
two morphological groups: early
(ellipticals + S0) and late (spirals + irregulars) types 
in the following tests:\\
(1) our galaxy sample is divided according to $\Sigma_{Gal}$ for bins in  $\Sigma_{Mass}$.
For each bin in $\Sigma_{Mass}$ we have divided our sample in three equal number sub-samples:
low, intermediate and high $\Sigma_{Gal}$. For the high and low projected galaxy density
sub-samples we computed  the fraction of early and late types as a
function of the global mass density (in the position of each object). The results of this
test are displayed in figures 4a and 4b.\\
(2) we have divided our galaxy sample according to $\Sigma_{Mass}$ for bins in  $\Sigma_{Gal}$. 
For each bin in $\Sigma_{Gal}$ we  divide our sample in three equal number sub-samples:
low, intermediate and high $\Sigma_{Mass}$. For the high and low projected mass density 
sub-samples we computed  the fraction of early and late types as a
function of the local galaxy density. The results of this second
 test are shown in figures 4c and 4d.\\

It should be recalled that the results shown in 
figures 4a and 4b show significant differences in the fraction of morphological types
between the high and low $\Sigma_{Gal}$ subsamples 
at low values of $\Sigma_{Mass}$, 
ie. typically in  the outskirts of the clusters.
On the other hand,
figures 4c and 4d show significant differences in the relative fraction of early and late types
between the high and low $\Sigma_{Mass}$ subsamples 
at high values of $\Sigma_{Gal}$, 
ie. in  the central virialized regions of clusters.
In order to provide a quantitative measurement of these effects 
we have computed the differences 
of the relative fraction of morphological types for the high and low density subsamples. 
We sum these differences and compute the corresponding 
averages and dispersions across the different bins
in total range of densities (see table 2, column 1).

Given the marked difference between the inner and outer
regions of the clusters, we have considered the associated threshold densities:
$\log(\Sigma_{Mass})\approx -0.03 $ and $\log(\Sigma_{Gal})\approx1.1$ (defining
the Low Density and High Density samples in table 2). 
These values correspond on average to a mean  overdensity $\delta_c  = 500$,
 a conservative estimate of the boundary between the inner, 
virialized region of the clusters and their recently accreted, still settling outer 
envelopes as discussed by  Evrard et al.(1996).
The results of tests 1 and 2 can be appreciated  by inspection to Table 2, where the average  
differences in morphological fractions, dispersions and statistical 
significance  as well as 
the percentages of galaxies used in each computation are listed.      
 
The results of tests 1 and 2 strongly suggest  
that local galaxy density should not be considered as the unique parameter that determines
the relative fractions of galaxy morphologies in clusters of galaxies. 
In the outskirts of clusters (quoted as Low Density in table 2), $\Sigma_{Gal}$  
accounts for most of the effect while $\Sigma_{Mass}$ may be considered 
a primary parameter in the high density virialized region of clusters.

\subsection{Gas density profile}

Other important global parameter worth to be considered is the intracluster gas density 
which may induce important morphological transformations in galaxies 
orbiting in the cluster potential well, such as ram pressure processes or gas 
evaporation.

For an isothermal gas distribution in hydrostatic equilibrium the  gas density profile 
can be obtained by fitting the X-ray surface brightness distribution with the well
known beta model:

(4) $$ I_x = I_0 [1+(r/r_c)^2]^{-3\beta + 1/2} $$

Fitting $\beta$ and $r_c$ from the previous equation the gas density can be derived as follows

(5) $$ \rho_{gas} = \rho_0 [1+(r/r_c)^2]^{-3\beta/2} $$

This equation can be projected in order to derive the projected intracluster 
gas density (Abadi and Navarro private communication):

(6) $$ \Sigma_{Gas}(r) = \rho_0 r_c [1+(r/r_c)^2]^{(1-3 \beta_f)/2} B(1/2,\frac{3\beta_f-1}{2}) $$

where $$ B(x,y) = \frac{\Gamma(x) \Gamma(y)}{\Gamma(x+y)}  $$

We have computed the fraction of galaxies by morphological 
types as a 
function of the $\Sigma_{Gas}$ for our sample. 
The results of this correlation are displayed in 
Figure 5. 

We have applied similar tests as those  described in the previous section with 
$\Sigma_{Gas}$ and $\Sigma_{Gal}$ so that now 
 we divide our galaxy sample according to $\Sigma_{Gal}$ for bins in  $\Sigma_{Gas}$
and for each bin in ($\Sigma_{Gas}$) we consider  three equal number sub-samples:
low, intermediate and high ($\Sigma_{Gal}$) and compute 
the fraction of morphological types of galaxies as a
function of the projected gas density for 
the high and low galaxy density 
sub-samples 
(test 3).
In a similar way we perform the complementary test for 
bins in  
$\Sigma_{Gal}$ and subsamples in $\Sigma_{Gas}$
(test 4). 

The results of these tests are displayed in figure 6
 and are also summarized in table 2. 

By inspection to the results of mass and gas projected profiles (figures 4 and 6, and 
table 2) vs. the local galaxy density we can infer that 
the relative fraction of galaxy morphology shows
a stronger correlation with the NFW profile in comparison to the $\beta$ model profile.

\subsection{Dependence on clustercentric distances normalized to $r_{opt}$ and $r_{500}$}

A usually adopted way to analyze the dependence of galaxy morphology on global 
cluster parameter is to compute galaxy clustercentric distances normalized to a
cluster characteristic radius.

WGJ have re-examined the relative fraction of galaxy morphological
types in Dressler's sample of  
clusters concluding that the clustercentric radial distance is a primary parameter 
in contrast to the local galaxy density, a secondary parameter in the morphology segregation. 
We have applied similar tests to those described in the previous
sections and for this case, our analysis is equivalent to that in WGJ.

We divide our galaxy sample according to $\Sigma_{Gal}$ for bins in $r/r_{opt}$ and 
for each bin in $r/r_{opt}$ we consider  three equal number sub-samples:
low, intermediate and high $\Sigma_{Gal}$ and compute 
the fraction of morphological types as a function of 
$r/r_{opt}$ for the high and low local galaxy density
sub-samples (test 5).
In a similar way we perform the complementary test for 
bins in  $\Sigma_{gal}$ and subsamples in
high and low normalized radial distance 
(test 6). 

The results of these tests are shown in 
Figure 7 and displayed in Table 2. It can be seen that neither 
$\Sigma_{Gal}$
nor $r/r_{opt}$ 
can be considered as a primary parameter defining the relation between galaxy morphology and environment.
The conclusion from a similar analysis by
WGJ  applied to the very central regions (within $r/R_{opt} = 0.25$, $\simeq 0.15 Mpc$ on average
for our sample)
is that a  normalized clustercentric distance 
$r/r_{opt}$ acts
as a single parameter driving the galaxy morphological segregation. 
This is consistent with our results
in the first two bins of figure 7a and 7b which correspond to these scales. 
By inspection to these figures, it can also  be appreciated the significant
dependence of the relative fraction of galaxy morphologies on 
$\Sigma_{gal}$ at larger distances from the cluster centers.

Several works use $r_{200}$ or $r_{500}$ as a characteristic cluster radius (see for instance
Yee et al. 1996) in order to make a proper comparison of different clusters. 
In this subsection we adopt the $r_{500}$ radius to normalize each galaxy radial distance
and we apply similar tests as those described previously.  
The results of the corresponding tests 7 and 8 are displayed in figure 8 (see also Table 2).
By inspection to these figures and Table 2 we conclude that
the dependence of galaxy morphology relative fractions 
on $r/r_{500}$ is similar to that obtained with
$r/r_{opt}$. Nevertheless, we find these dependencies
considerably less significant than that  on the
NFW projected mass density profile as it can be seen by comparison of figures 4 and 8.

\subsection{Morphological segregation in clusters with low / high X-ray luminosity and 
intracluster gas temperature}

Different authors have 
analyze possible dependencies of the morphological segregation on the X-ray 
luminosity. Dressler 1980b computed the T-$\Sigma$ relationship for  eight 
strong X-ray emitters  ($\L_x \geq 10^{44} erg s^{-1}$)  finding 
similar relations between morphological types and galaxy density than in the total sample.
WGJ divided a sample of 39 clusters with X-ray 
luminosity in three sub-samples 
finding no significant dependence of the morphological fractions as a function of 
clustercentric radius on the cluster 
X-ray luminosity.
For our sample 
X-ray luminosities and intracluster gas temperatures 
are available for all clusters. 
In order to explore the correlations between the fraction of morphological types
and the global and local parameters in different cluster environments
we have divided our sample into high and low cluster temperature and luminosity 
subsamples.
In a similar way as done in the previous sections, 
we computed the fraction of spiral galaxies
for high and low $\Sigma_{Mass}$ and $\Sigma_{Gal}$ values in bins
of  $\Sigma_{Gal}$ and $\Sigma_{Mass}$ respectively, for two subsamples
of clusters, luminous ($L_X \geq 1.638\cdot 10^{44} erg.s^{-1} $) and 
non luminous ($ L_X \leq 1.123\cdot 10^{44} erg.s^{-1} $).
We have also applied the same analysis to the two subsamples
defined by hot ($T_X \geq 3.7 keV.$), and cold ($T_X \leq 3.4 keV.$) clusters.
The results of these tests show 
a lack of a strong dependence on luminosity and temperature, in agreement with Dressler 1980b
and WGJ. However, 
in the central regions of the subsample of high temperature and X-ray luminosity clusters 
we notice a stronger dependence of the relative morphological fractions 
on the global mass density vs. the local galaxy density when compared to low 
temperature/X-ray luminosity cluster subsample.


\section{DISCUSSION AND CONCLUSIONS}

In hierarchical scenarios for 
structure formation in the universe, clusters are assembled from smaller
subunits, so that the local galaxy density, associated to these primordial
clumps, could play a significant role in the segregation of morphological types. 
On the other hand, the several mechanisms related to galaxy
evolution within the potential well of clusters
(RAM pressure, tidal effects, etc.) could also explain  
the morphological segregation and its relation to global parameters.

The results of our analysis suggests that there are different mechanisms controlling 
the morphological segregation depending  on  the galaxy  environment.
We find that mechanisms of global nature dominate in high density environments,
namely, the virialized region of clusters,  while local  
galaxy density as defined by Dressler 1980b is the relevant parameter
in the outskirts where the influence of the cluster as a whole is
relatively small compared to local effects.
 
As it can be inferred by inspection to figure 4, a primary parameter in 
the segregation of morphological types in high density regions is the
global cluster mass density at the position of the galaxies,
computed using the scaling relationship between the mean cluster 
temperature  and the projected  NFW mass density profile.  
We find that the relative fraction of galaxy morphologies 
shows a stronger dependence on the local NFW projected density profile 
compared to other radial distances normalizations such as $r_{500}$ or $r_{opt}$.
Therefore, these results might be applied to other studies of 
galaxy properties in clusters
such as star formation rate, fraction of blue galaxies, etc. that show a significant
dependence on radial distance to the cluster centers. 

By comparison of the results corresponding to the NFW and the $\beta$ model
 profiles (projected mass and gas density respectively) we conclude that
the relative fraction of galaxy morphologies is more strongly correlated with
the NFW profile.  
This result may serve to assess 
the importance of tidal effects and gaseous phenomena operating
in the transformation of spiral galaxies into S0s.

Our results give support to the idea that  
tidal force effects produced by the cluster potential, galaxy harassment, or truncated 
star formation among others physical mechanisms
would be primary in driving the observed
morphological segregation  in clusters of galaxies. 

We have also  taken into account 
different cluster properties in our analysis 
by considering subsamples of high/low cluster temperature 
and high/low X-ray luminosity.
The results of these analysis
are  similar to those obtained for the total sample although
in the high density regions of hot clusters we find a tendency of a stronger
morphological segregation as a function of  
$\Sigma_{Mass}$.


\section{Appendix: Corrections for background / foreground galaxies and magnitude 
cutoff.}

As is was extensively discussed by WGJ,  corrections due to background/foreground
galaxies are to be applied  when dealing with relative morphological fractions
in clusters. For those computations requiring estimates of the local
galaxy density we have applied corrections in the same way as WGJ.  
We recall that the total mass 
density  profile is obtained from the mean cluster temperatures 
of the intracluster gas, estimates which are nearly free of projection effects. 

However, estimates of the relative fraction of galaxy 
morphological types will be biased when a background/foreground galaxy is taken 
as a center to compute the corresponding  gas/mass density. To correct for  this 
effect we have  take into account the correlation between  galaxy projected density and
the percentage of background/foreground galaxies given by WGJ. We estimate the corrections
assuming a correlation between mass/gas density and the local galaxy density (see
Figure 3 which is a good approximation given 
the small difference of observed and actual 
relative fraction of morphological types due to projection effects.
Using the correlation found by WGJ between absolute magnitude 
and cumulative number of galaxies by morphological types we have also corrected   
for absolute magnitude cutoff. This effect takes into account the fact that clusters 
are at different distances with the same limiting apparent magnitude.


\noindent Acknowledgments     

This work was partially supported by the Consejo de Investigaciones Cient\'{\i}ficas y 
T\'ecnicas de la Republica Argentina, CONICET, the Consejo de Investigaciones 
Cient\'{\i}ficas y Tecnol\'ogicas de la Provincia de C\'ordoba, CONICOR, and Fundaci\'on 
Antorchas, Argentina. We thank the Referee for helpful suggestions.

\begin{deluxetable}{crclrclccccccc}
\pagestyle{empty}
\tablenum{1}
\tablewidth{17cm}
\tablecaption{Cluster Sample.}
\tablehead{ Name & \multicolumn{3}{c}{RA} & \multicolumn{3}{c}{Dec.} & z & $T_{x}$
& $r_c$ & $\beta$ & $\rho_0$ & $r_{opt}$ & $r_{500}$   \nl
 & \multicolumn{3}{c}{[1950]}  & \multicolumn{3}{c}{[1950]}  & &  keV & Mpc & & $10^{-3}\rm cm^{-3}$ &  Mpc & Mpc}
\startdata
  A0076  &   00 & 37 & 25.1 & +06 & 33 & 32 & 0.0416  &  1.5 & 0.41 &  0.60 &  0.522 & 0.23  & 0.90  \nl
  A0119  &   00 & 53 & 43.5 & -01 & 31 & 28 & 0.0440  &  5.9 & 0.32 &  0.53 &  1.221 & 0.83  & 1.79  \nl
  A0154  &   01 & 08 & 22.2 & +17 & 23 & 37 & 0.0658  &  3.1 & 0.17 &  0.55 &  1.690 & 0.80  & 1.25  \nl
  A0194  &   01 & 23 & 20.0 & -01 & 38 & 12 & 0.0178  &  1.4 & 0.20 &  0.60 &  0.719 & 0.38  & 0.90  \nl
  A0376  &   02 & 42 & 57.3 & +36 & 41 & 52 & 0.0488  &  5.1 & 0.08 &  0.46 &  3.941 & 1.05  & 1.65  \nl
  A0400  &   02 & 55 & 00.0 & +05 & 48 & 25 & 0.0232  &  2.5 & 0.26 &  0.65 &  4.430 & 0.43  & 1.20  \nl
  A0496  &   04 & 31 & 20.4 & -13 & 21 & 48 & 0.0320  &  3.9 & 0.14 &  0.59 &  5.283 & 0.39  & 1.48  \nl
  A0539  &   05 & 13 & 55.2 & +06 & 23 & 16 & 0.0205  &  3.0 & 0.11 &  0.60 &  3.054 & 0.56  & 1.32  \nl
  A0592  &   07 & 39 & 56.5 & +09 & 29 & 30 & 0.0624  &  3.2 & 0.14 &  0.60 &  2.606 & 0.26  & 1.28  \nl
  A0957  &   10 & 11 & 07.9 & -00 & 40 & 53 & 0.0440  &  2.8 & 0.14 &  0.52 &  1.720 & 0.56  & 1.23  \nl
  A1142  &   10 & 58 & 17.7 & +10 & 47 & 40 & 0.0353  &  3.7 & 0.19 &  0.60 &  0.786 & 0.28  & 1.43  \nl
  A1185  &   11 & 08 & 03.0 & +28 & 59 & 04 & 0.0304  &  3.9 & 0.15 &  0.62 &  1.445 & 0.39  & 1.48  \nl
  A1377  &   11 & 44 & 40.6 & +55 & 59 & 40 & 0.0509  &  2.7 & 0.29 &  0.60 &  0.793 & 0.42  & 1.20  \nl
  A1656  &   12 & 57 & 18.3 & +28 & 12 & 22 & 0.0235  &  8.1 & 0.43 &  0.67 &  2.275 & 1.24  & 2.16  \nl
  A1913  &   14 & 24 & 25.5 & +16 & 53 & 40 & 0.0533  &  2.9 & 0.57 &  0.60 &  0.396 & 0.61  & 1.24  \nl
  A1983  &   14 & 50 & 36.8 & +16 & 55 & 02 & 0.0458  &  2.2 & 0.08 &  0.60 &  3.838 & 0.60  & 1.09  \nl
  A1991  &   14 & 52 & 13.4 & +18 & 50 & 56 & 0.0586  &  5.4 & 0.06 &  0.56 & 10.810 & 0.25  & 1.67  \nl
  A2040  &   15 & 10 & 21.0 & +07 & 37 & 06 & 0.0456  &  2.5 & 0.14 &  0.60 &  1.782 & 0.48  & 1.16  \nl
  A2256  &   17 & 06 & 44.3 & +78 & 42 & 46 & 0.0601  &  7.5 & 0.58 &  0.76 &  1.278 & 1.05  & 1.97  \nl
  A2634  &   23 & 35 & 54.9 & +26 & 44 & 19 & 0.0312  &  3.4 & 0.42 &  0.60 &  0.597 & 0.98  & 1.38  \nl
  A2657  &   23 & 42 & 22.9 & +08 & 54 & 15 & 0.0414  &  3.4 & 0.14 &  0.52 &  3.018 & 0.73  & 1.36  \nl
  Cent.  &   12 & 46 & 03.4 & -41 & 02 & 26 & 0.0107  &  3.9 & 0.15 &  0.45 &  1.939 & ----  & 1.52
\enddata
\end{deluxetable}

\begin{deluxetable}{lccc}
\pagestyle{empty}
\tablenum{2}
\tablewidth{14cm}
\tablecaption{Results}

\tablehead{     &        Total       &     Outskirts     &      Inner  }
\startdata
Test 1   &   0.12 $\pm$ 0.03 [4.6]  &  0.20 $\pm$ 0.06 [3.3] &  0.07 $\pm$ 0.02 [3.1] \nl
                &                    &       37\%        &       63\%        \nl
Test 2   &   0.09 $\pm$ 0.03 [3.0] & -0.01 $\pm$ 0.07 [0.2] &  0.16 $\pm$ 0.03 [5.3]  \nl
                &                    &       40\%        &       60\%        \nl
\hline
Test 3  &   0.12 $\pm$ 0.03 [4.4] &  0.07 $\pm$ 0.04 [1.9] &  0.13 $\pm$ 0.03 [4.8]  \nl
                &                    &       37\%        &       63\%         \nl
Test 4  &   0.06 $\pm$ 0.03 [2.0] & -0.01 $\pm$ 0.07 [0.1] &  0.11 $\pm$ 0.03 [5.1]  \nl
                &                    &       49\%        &       51\%         \nl
\hline
Test 5  &    0.16 $\pm$ 0.04 [4.3] &  0.24 $\pm$ 0.07 [3.2] &  0.10 $\pm$ 0.03 [2.9]  \nl
                &                    &       41\%        &       59\%         \nl
Test 6  &   -0.09 $\pm$ 0.03 [3.0] & -0.10 $\pm$ 0.07 [1.4] & -0.08 $\pm$ 0.02 [3.4]  \nl
                &                    &       47\%         &      53\%              \nl
\hline
Test 7  &    0.12 $\pm$ 0.03 [4.0] & 0.13 $\pm$ 0.06 [2.1] & 0.10 $\pm$ 0.03 [3.1]  \nl
                &                    &    45\%            &      55\%          \nl
Test 8  &    -0.04 $\pm$ 0.03 [1.0] & 0.03 $\pm$ 0.08 [0.4] & -0.08 $\pm$ 0.03 [2.6] \nl
                &                    &    42\%            &     58\%         \nl
\enddata
\end{deluxetable}

\clearpage

\figcaption{Figure 1. Relative fraction of E (solid line), 
S0 (dotted line) and S+I (dashed line) as a function of the local galaxy 
density.}

\figcaption{Figure 2. Relative fraction of E, S0 and S+I as a function of the projected 
mass density. Line types are the same as in figure 1.}

\figcaption{Figure 3. Correlation between local galaxy density $\Sigma_{Gal}$ and 
projected cluster mass density $\Sigma_{Gas}$ at the position of each galaxy of the cluster sample.
The solid line correspond to the best power-law fit.}

\figcaption{Figure 4.  Relative fraction of galaxy morphological types as a function of the global 
projected mass density $\Sigma_{Mass}$ and local galaxy density $\Sigma_{Gal}$.
a) Fraction of E+S0 vs. $\Sigma_{Mass}$ at high (solid line),
 and low (dashed line) $\Sigma_{Gal}$.
b) Same as 4a for the relative fraction of S+I.
c) Fraction of E+S0 vs. $\Sigma_{Gal}$ at high (solid line),
 and low (dashed line) $\Sigma_{Mass}$.
d) Same as 4c for the relative fraction of S+I.}

\figcaption{Figure 5. Relative fraction of E, S0 and S+I as a function of the projected 
gas density. Line types are the same as in figure 1.}

\figcaption{Figure 6.  Relative fraction of galaxy morphological types as a function of the global 
projected gas density $\Sigma_{Gas}$ and local galaxy density $\Sigma_{Gal}$.
a) Fraction of E+S0 vs. $\Sigma_{Gas}$ at high (solid line),
 and low (dashed line) $\Sigma_{Gal}$.
b) Same as 6a for the relative fraction of S+I.
c) Fraction of E+S0 vs. $\Sigma_{Gal}$ at high (solid line),
 and low (dashed line) $\Sigma_{Gas}$.
d) Same as 6c for the relative fraction of S+I.}

\figcaption{Figure 7.  Relative fraction of galaxy morphological types as a function of 
the clustercentric projected radial distance normalized to $r_{opt}$
and local galaxy density $\Sigma_{Gal}$.
a) Fraction of E+S0 vs. $r/r_{opt}$ at high (solid line),
 and low (dashed line) $\Sigma_{Gal}$.
b) Same as 7a for the relative fraction of S+I.
c) Fraction of E+S0 vs. $\Sigma_{Gal}$ at high (solid line),
 and low (dashed line) $r/r_{opt}$.
d) Same as 7c for the relative fraction of S+I.}

\figcaption{Figure 8.  Relative fraction of galaxy morphological types as a function of 
the clustercentric projected radial distance normalized to $r_{500}$
and local galaxy density $\Sigma_{Gal}$.
a) Fraction of E+S0 vs. $r/r_{500}$ at high (solid line),
 and low (dashed line) $\Sigma_{Gal}$.
b) Same as 8a for the relative fraction of S+I.
c) Fraction of E+S0 vs. $\Sigma_{Gal}$ at high (solid line),
 and low (dashed line) $r/r_{500}$.
d) Same as 8c for the relative fraction of S+I.}


\begin{references}

\reference{} Abadi, M. G., Moore, B., and Bower, R. G. 1999, MNRAS 308, 947.
\reference{} Barrow, J. D., Sonoda, D. H., Bhavsar, S. P. 1984, MNRAS 210, 19. 
\reference{} Bartelmann, M., 1996 Astron. Astrphys. 313, 697.
\reference{} Byrd, G., and Valtonene, M., 1990 ApJ. 350, 89.
\reference{} Cowie,L. L., and Songaila,A. 1977 Nature 266, 501.
\reference{} Dressler, A. 1980a, ApJS. 424,565.
\reference{} Dressler, A. 1980b, ApJ. 236, 351.
\reference{} Dressler, A., Oemler, A. Jr., Cousch, W. J., Smail, I., Ellis, R. S., Barger, A.,
Butcher, H., Poggianti, B. M., and Sharples, R. M. 1997, ApJ. 490, 577 (D97).
\reference{} Evrard A. E. , Metzler, C. A., and Navarro, J. F. 1996 ApJ. 469, 494.
\reference{} Gunn, J. E., and Gott, J. R. 1972, ApJ. 176, 1.
\reference{} Jones, C., and Forman, W. 1999 ApJ. 511,.65.
\reference{} Lavery, R. J., and Henry, P. J. 1988, ApJ 330, 596.
\reference{} Larson, R. B., Tinsley, B. M., and Caldwell, c. n. 1980 ApJ. 237, 692.
\reference{} Melnick, J., and Sargent, W. L.1997, ApJ. 215, 401.
\reference{} Miller, R. H., 1988 Comments on Ap. 13, 1.
\reference{} Moore, B.,Katz, N., Lake, G., Dresssler, A., and Oemler, A., 1996 Nature 379, 613.
\reference{} Navarro, J. F., Frenck, C. S., and White, S. D. M. 1995 MNRAS 275, 720 (NFW).
\reference{} Oemler A. Jr. 1974, ApJ. 194, 1.
\reference{} Ostriker, J. P., and Tremaine, S. D., 1975 ApJ (letters) 202, L113.
\reference{} Samrom$\grave{a}$, M., and Salvador-Sole, E., 1990, ApJ. 360, 16.
\reference{} Valotto, C., Moore, B., and Lambas, D. 2001, ApJ, in press.
\reference{} Whitmore, B. C., and Gilmore, D. M. 1991, ApJ. 367, 64.
\reference{} Whitmore, B. C., Gilmore, D. M., and Jones, C. 1993, ApJ. 407, 489 (WGJ).

\end{references}
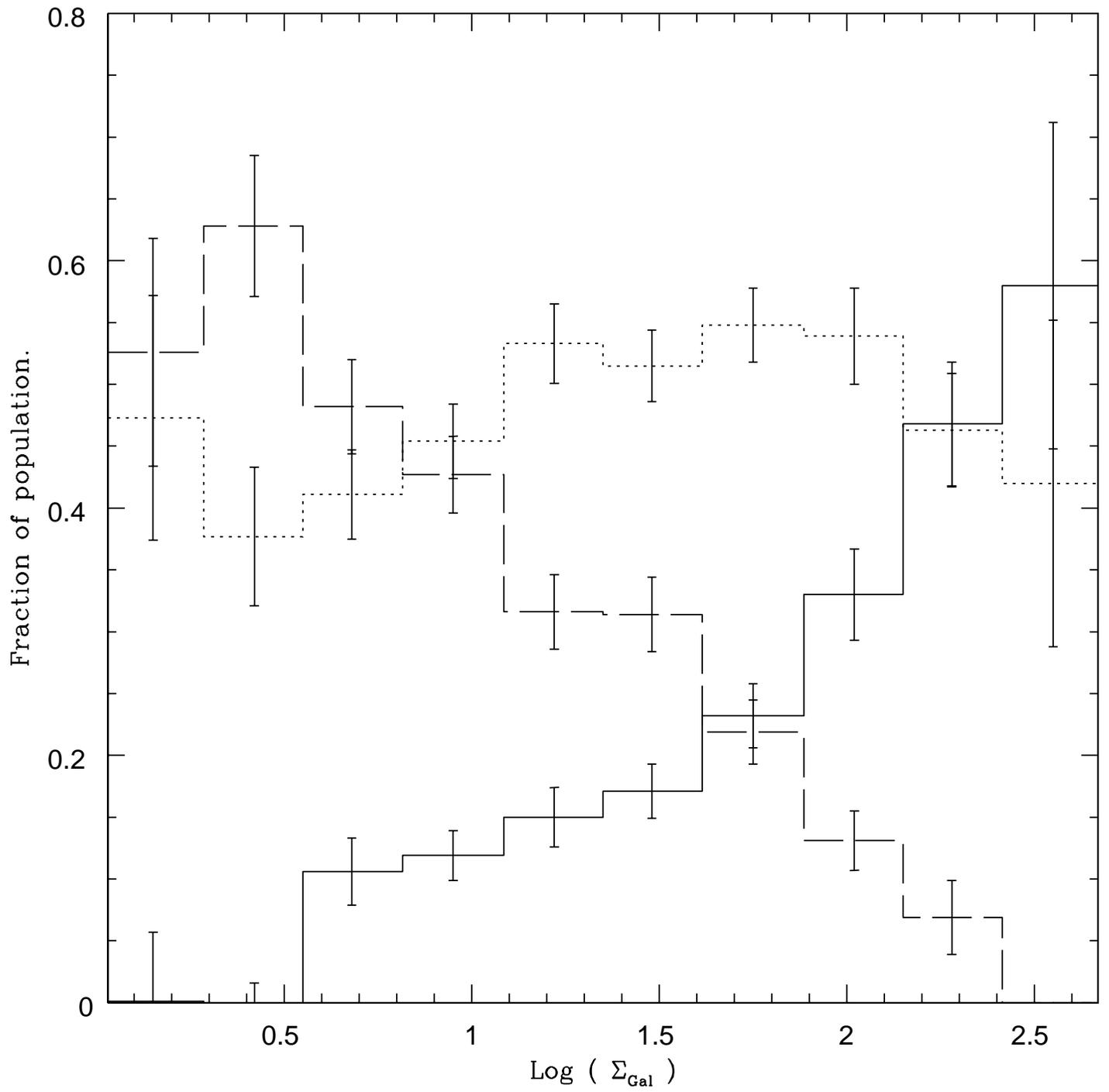
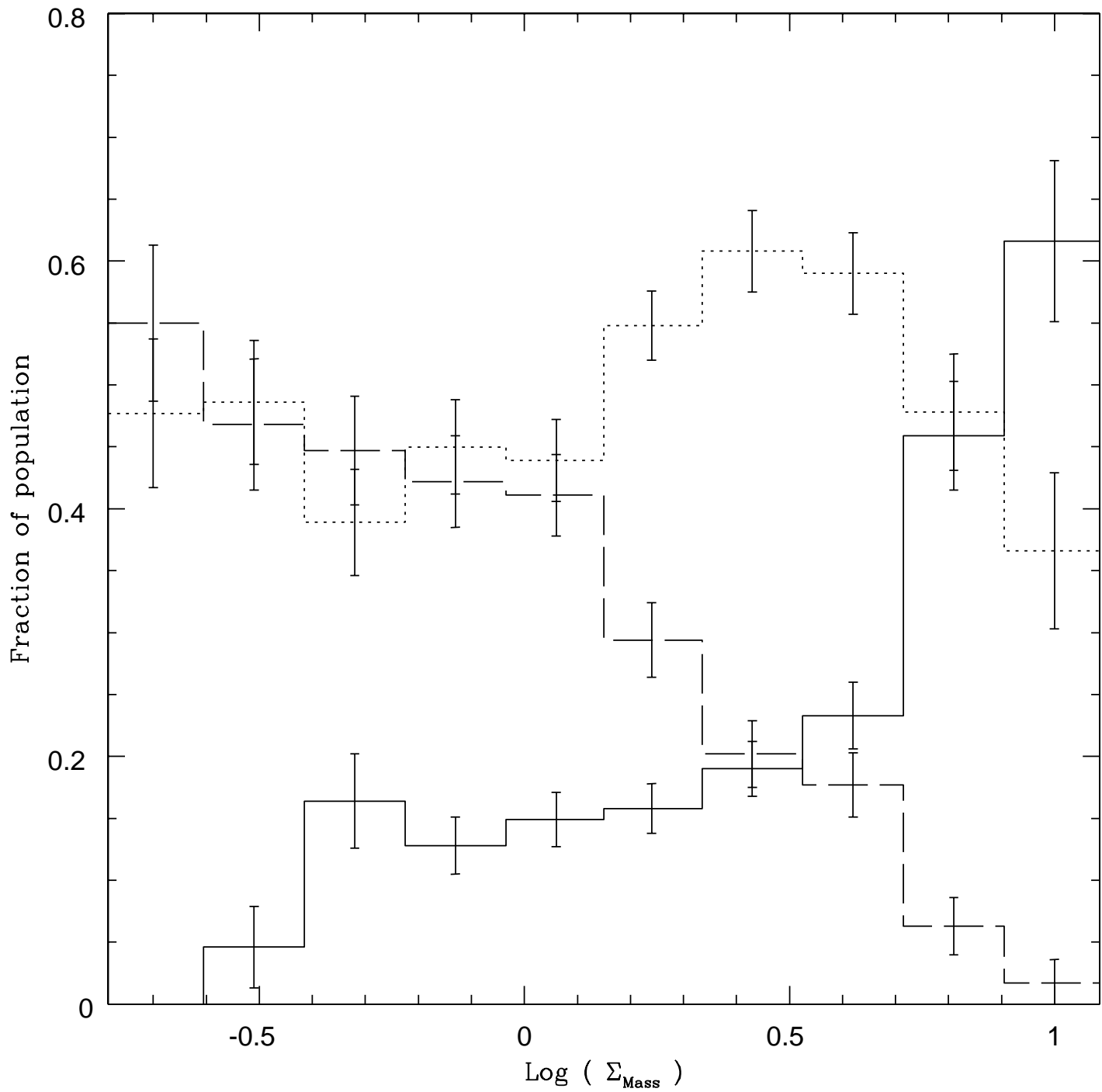
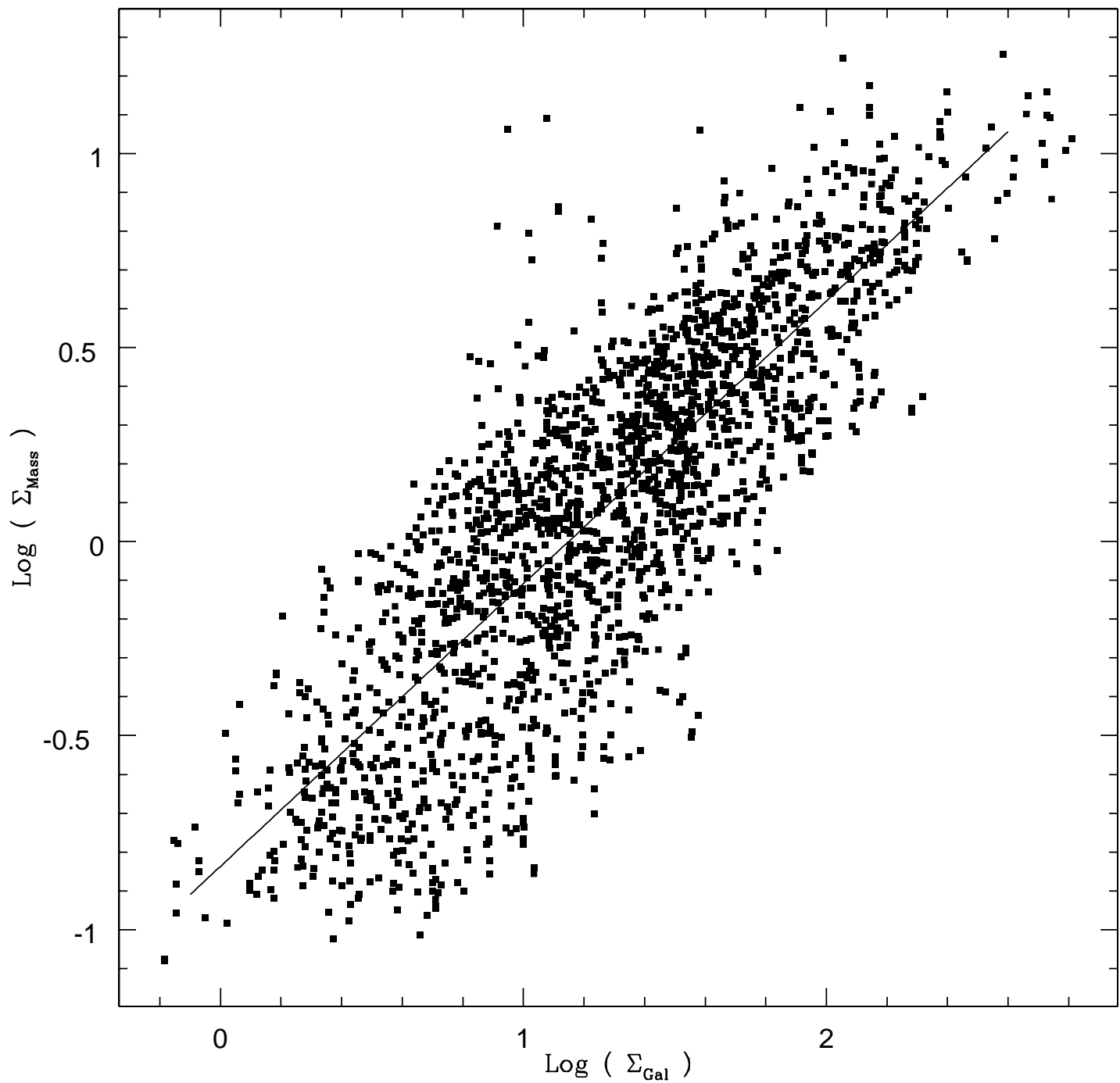
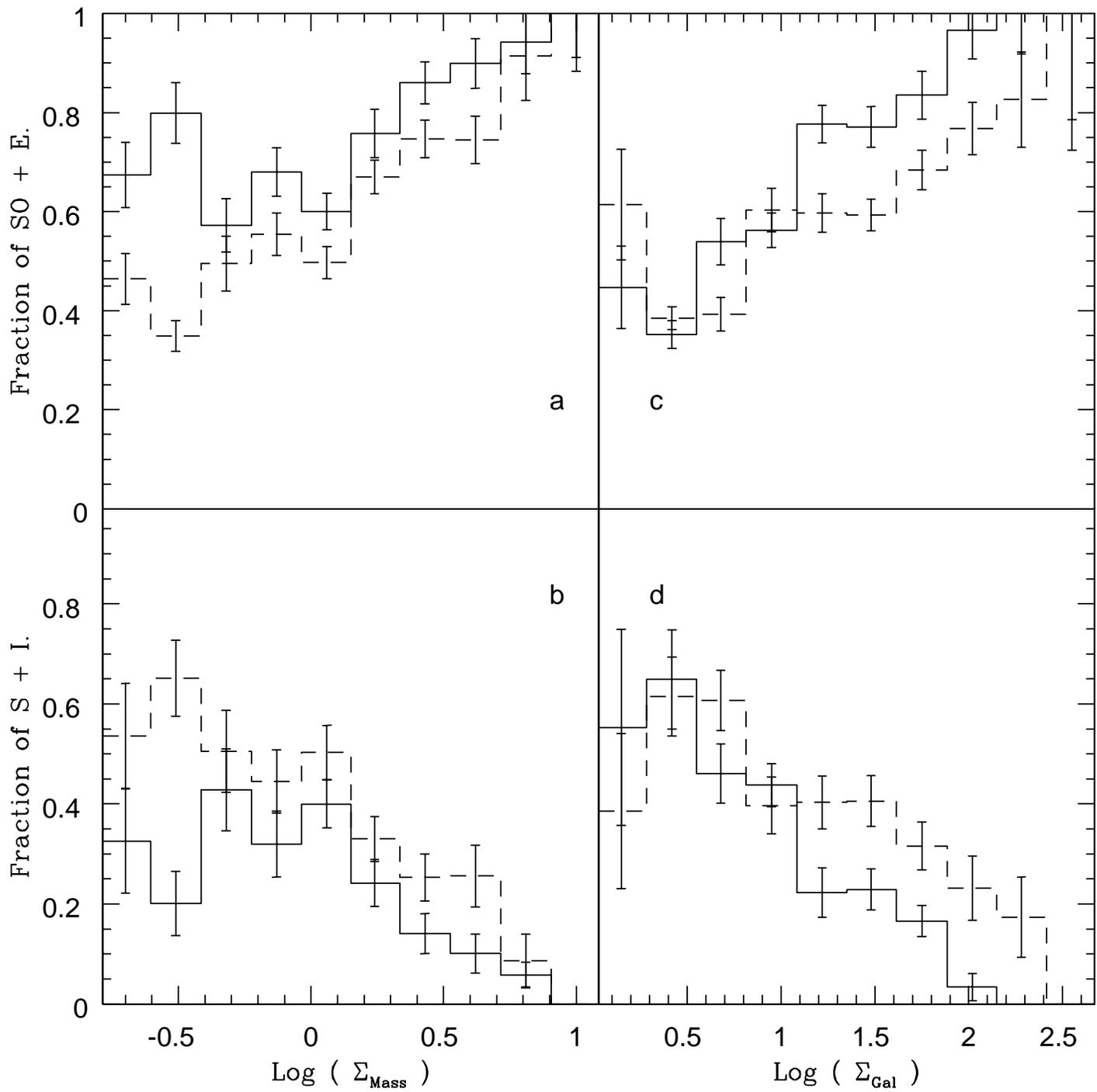
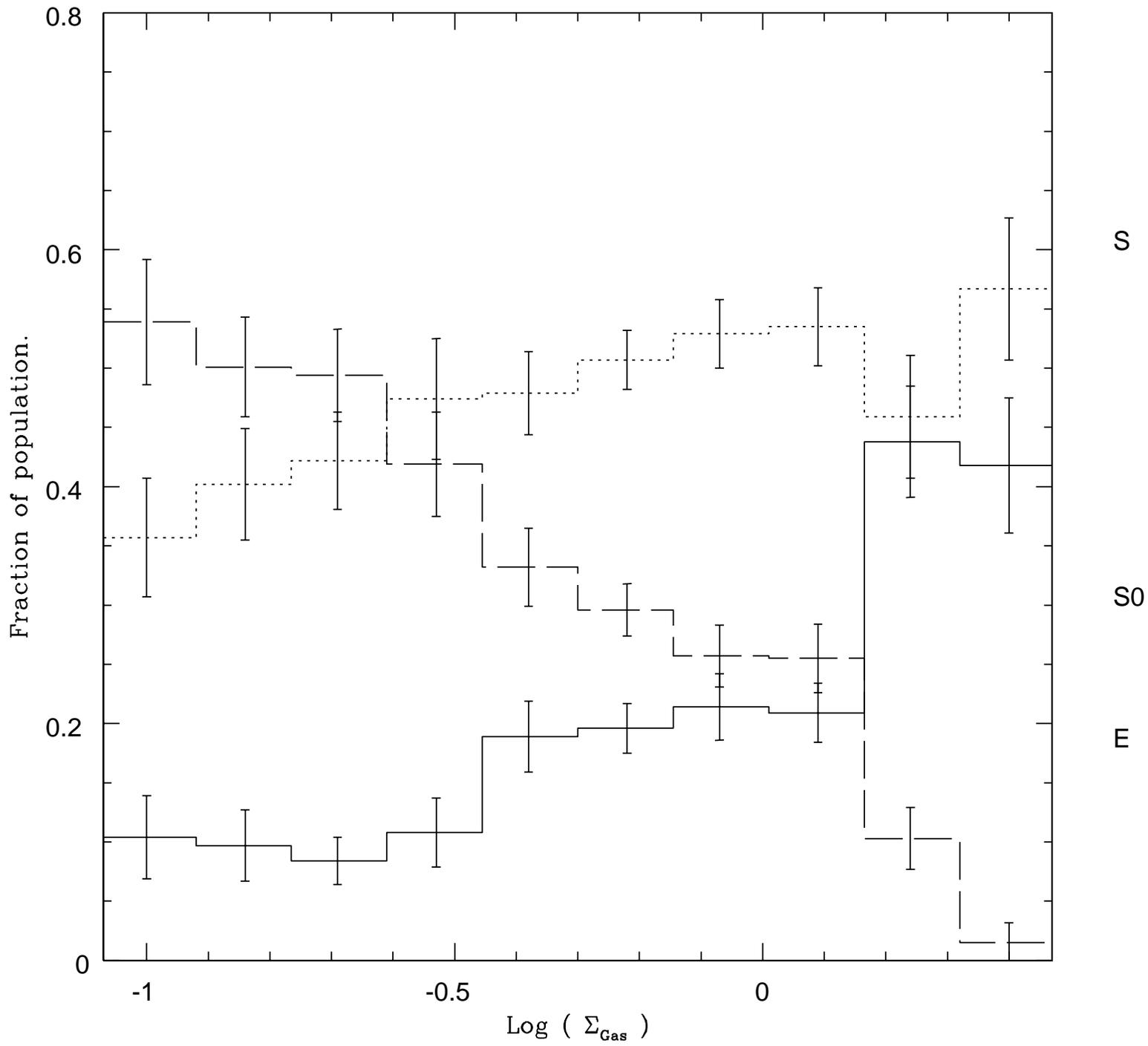
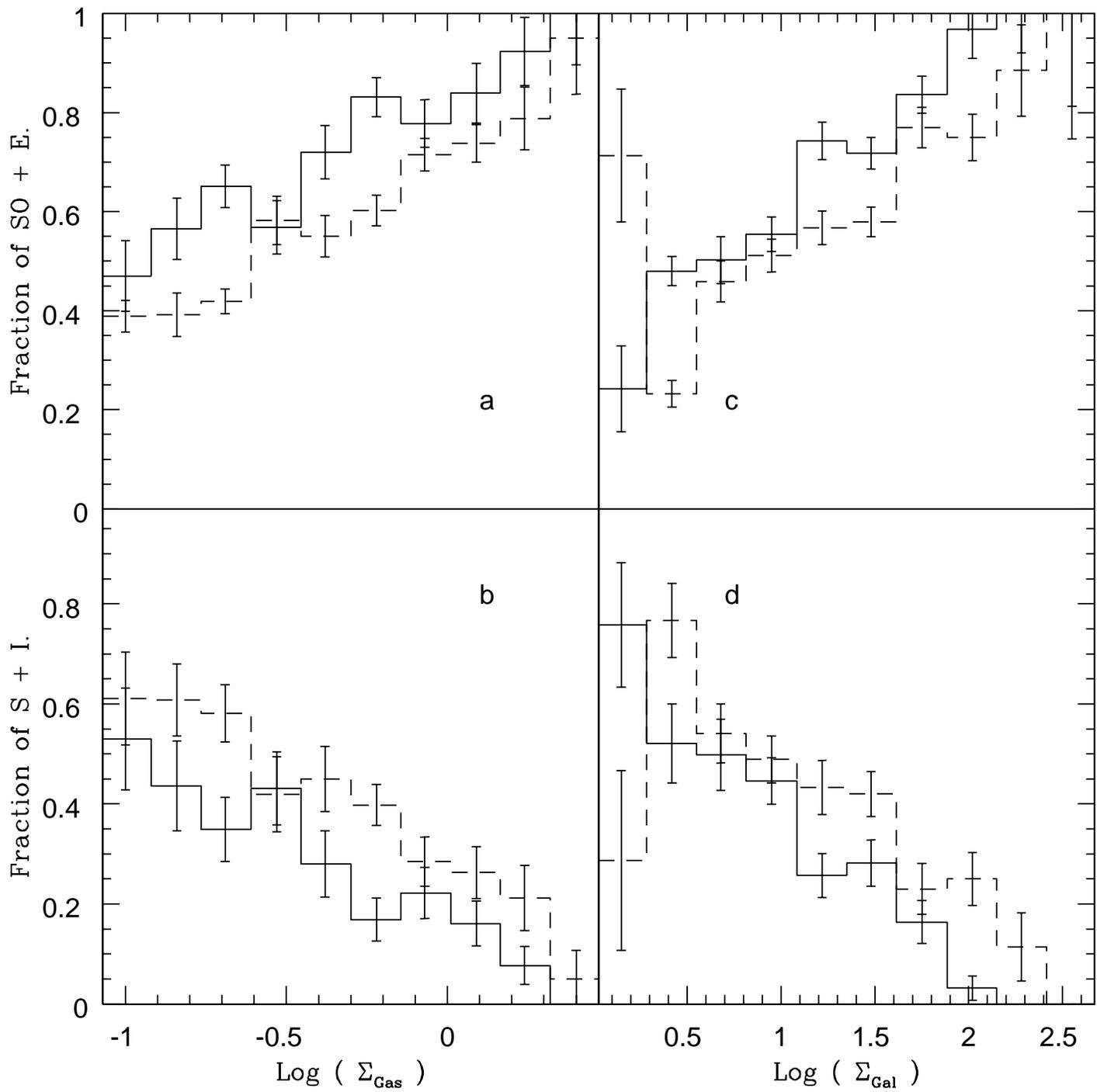
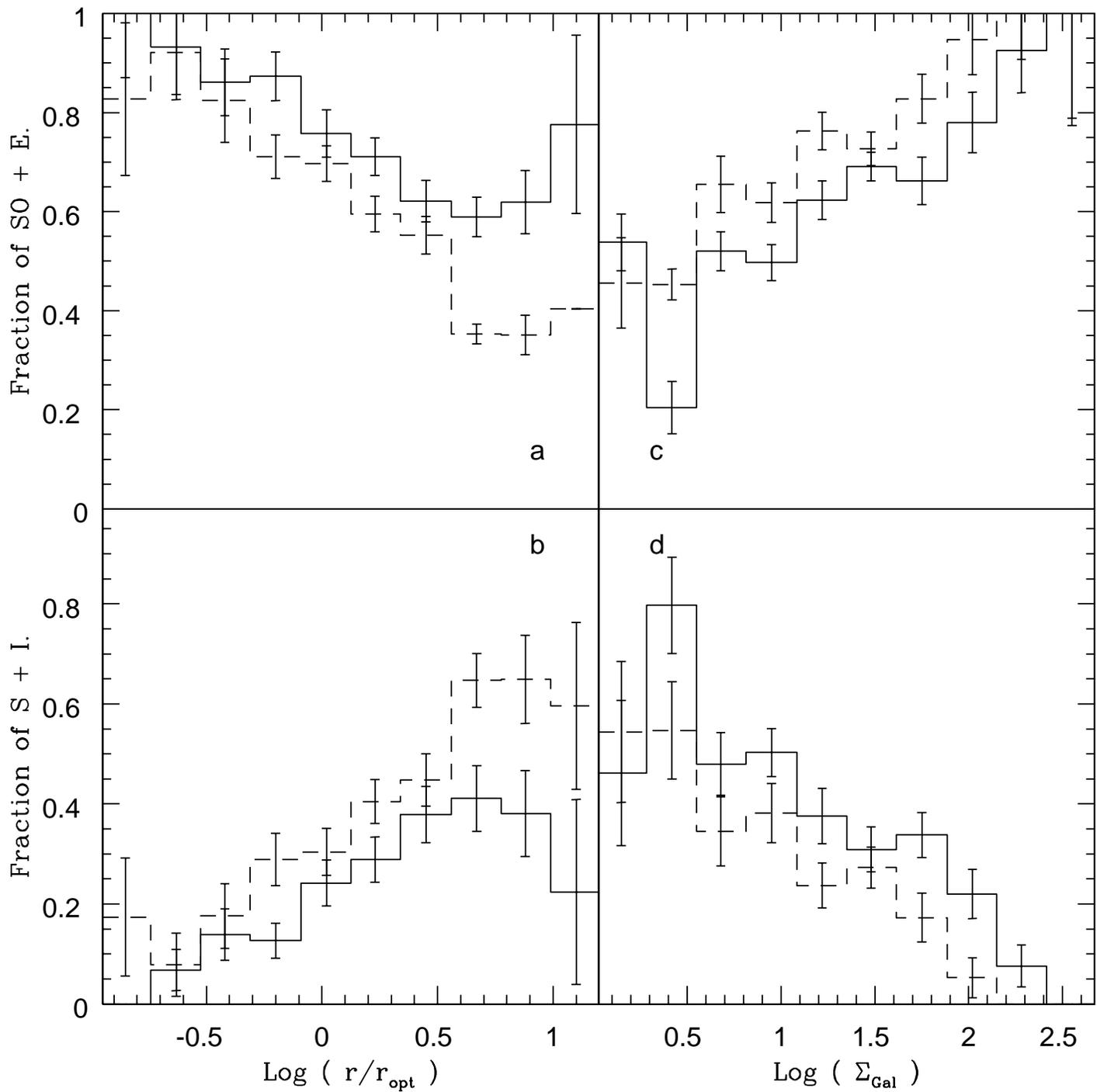
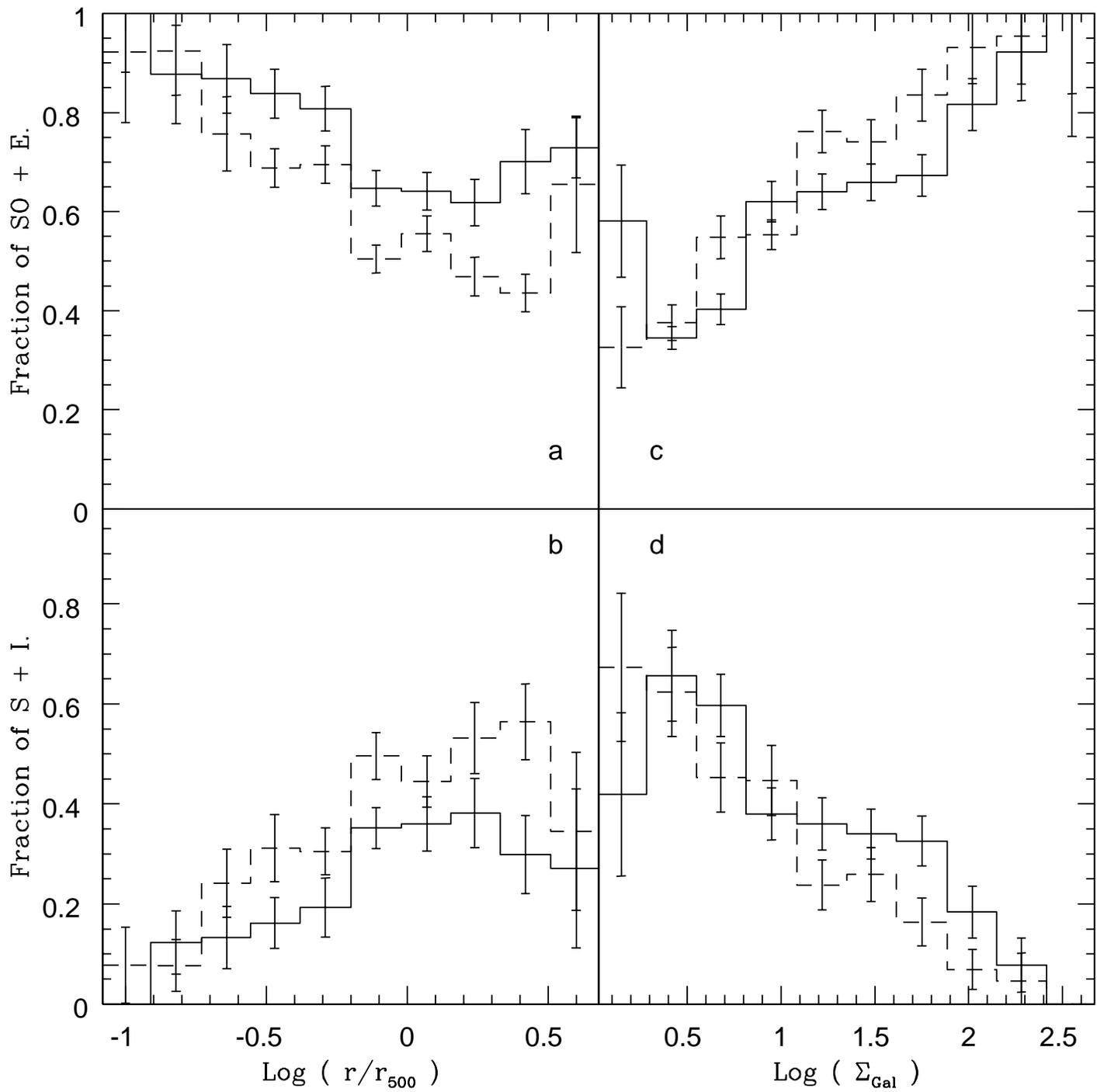
\end{document}